\documentstyle[multicol,aps,epsfig]{revtex}

\begin{document}

\draft 

\title{Observation of Partially Suppressed Shot Noise in Hopping
Conduction}

\author{ V. V. Kuznetsov and E. E. Mendez } 

\address{ Department of Physics and Astronomy, State University of New
York at Stony Brook, Stony Brook, NY 11794-3800 }

\author{E. T. Croke, X. Zuo, and G. L. Snider }

\address{ Department of Electrical Engineering, University of Notre
Dame, Notre Dame, IN 46556 } 

\maketitle

\begin{abstract}

We have observed shot noise in the hopping conduction of two
dimensional carriers confined in a p-type SiGe quantum well at a
temperature of 4K. Moreover, shot noise is suppressed relative to its
``classical'' value 2eI by an amount that depends on the length of the
sample and carrier density, which was controlled by a gate voltage. We
have found a suppression factor to the classical value of about one
half for a 2 $\mu$m long sample, and of one fifth for a 5 $\mu$m
sample. In each case, the factor decreased slightly as the density
increased toward the insulator-metal transition.  We explain these
results in terms of the characteristic length ($\simeq 1\mu$m in our
case) of the inherent inhomogeneity of hopping transport.

\end{abstract}

\begin{multicols}{2}

Shot noise, which is a manifestation of the particle nature of the
electric current, has lately received much attention \cite{review}
because it can yield information complementary to that obtained from
conductance measurements.  It is most pronounced when the current is
formed by statistically independent charges tunneling through a single
potential barrier of low transparency, in which case the noise power
spectral density, $S$, is equal to the Schottky, or classical, value
of $2qI$, where $q$ is the value of the charge and $I$ the average
current. This proportionality has been employed, for instance, to
determine the effective charge in superconducting transport \cite{ns}
and in the fractional quantum Hall effect \cite{fqhe}. In more general
cases, when the motion of the charges is not independent from each
other, the value of $S$ has an additional factor $F$, the so called
Fano factor.  Except when negative differential conductance occurs
\cite{mine1}, the Fano factor is in the range $0<F<1$, meaning that
shot noise is then partially suppressed.  Knowledge of the degree of
suppression sheds light at the microscopic level on the conduction
mechanisms of a specific system.

By now the noise characteristics of ballistic, diffusive, and chaotic
transport have been established, as well as those of resonant
tunneling and single-electron tunneling.  For example, in diffusive
conductors, whose size along the current direction is smaller than the
inelastic scattering length, it is $F=1/3$.  In the opposite limit,
that is, when the scattering length is much shorter than the length of
the sample, $F$ approaches zero, and in macroscopic metallic
conductors shot noise is completely suppressed.

Although the 1/f noise properties of hopping conduction have also been
elucidated \cite{kogan}, surprisingly, little is known about shot
noise for such a well studied transport mechanism \cite{hopping},
which has regained interest in connection with the metal-insulator
transition recently observed in Si MOSFETs \cite{mosfet} and other
two-dimensional (2D) systems, including SiGe quantum wells
\cite{sige}.  In this Letter we report the observation in a 2D hopping
conductor of shot noise that is only partially suppressed, and we
introduce a model that in spite of its simplicity can explain our
experimental results.

If in hopping conduction, where electrons tunnel assisted by phonons
between localized states created by the random impurity potential, the
determinant factor were the inelastic scattering length, then, given
the smallness of this length, shot noise should be zero.  On the other
hand, since the process involves tunneling through potential barriers,
which insures the discrete nature of the current, one could naively
assume that shot noise should have the full $2qI$ value.  A closer
look reveals a more complex situation.

In a simple one-dimensional system in which electrons tunnel through N
identical barriers, the Fano factor is $F=1/N$.  When, like in
hopping, tunneling occurs between single electron states, depending on
their occupancy, shot noise suppression can be a different function of
$N$ \cite{korotkov}.  Since the equivalent resistances of the various
hops are exponentially different from each other and only the most
resistive hop (``bottle neck'') determines the current, it could be
argued that effectively $N=1$.  However, in real quasi one-dimensional
hopping \cite{gershenson}, in which there is a maximum resistance
obtainable (hard hop), the effective N should be the number of hard
hops along the sample length, as in the case of identical barriers.

In a 2D system, hopping conduction can be seen as occurring through a
network of one-dimensional chains connected to each other at certain
nodes, as is normally done in percolation theory, where the network is
modeled by resistors of exponentially different values, out of which
the most conductive subnetwork (critical percolation subnetwork) is
selected \cite{hopping}. The characteristic size of this subnetwork is
the length beyond which the sample is homogeneous, and its nodes are
such that each chain contains only one resistor with the largest
resistance (hard hop).  Even this simpler subnetwork is still
complicated enough as to make it difficult to guess, let alone to
calculate, what the effective $F$ will be.

To answer experimentally this question we chose a 2D hole system
confined in a modulation-doped SiGe well.  The heterostructure, grown
by molecular beam epitaxy on a n-Si substrate, consists (from the
substrate up) of 4300\AA\ of Si boron-doped at $1\times10^{18}$
$cm^{-3}$, 225\AA\ of undoped Si, 500\AA\ of Si$_{0.8}$Ge$_{0.2}$
(quantum well), 275\AA\ of undoped Si, and finally 725\AA\ of Si
boron-doped at $1\times10^{18}$ $cm^{-3}$. The 2D hole density and the
in-plane resistivity were controlled by a voltage $Vg$ applied to an
Al Schottky gate deposited on the top layer.  The heterostructure was
processed into samples with gate width of 50$\mu$m and length (along
the current direction) of either 2$\mu$m or 5$\mu$m.  The noise
measurements were done with the samples immersed in liquid He.  At
$T=4K$ even for the smallest $Vg$ used the samples had resistance per
square much larger than the quantum resistance $h/2e^2$, so that they
were always in the insulating regime.

Current through the sample was produced by applying a $dc$ bias and a
small $ac$ signal, $Vin$, to a $1M \Omega$ load resistor.  The voltage
drop across the sample, $Vsd$, was measured simultaneously with the
$ac$ signal and the noise, using a lock-in technique and a spectrum
analyzer.  The $1/f$-noise contribution was reduced by doing the
measurements at high frequencies up to 100kHz, which demanded
minimizing lead capacitance and required placing a preamplifier inside
the cryostat and very close to the sample. The preamplifier, with an
output resistance of about 100$\Omega$, was a commercial low power
switching MOSFET, connected in source follower configuration, as in
\cite{birk}.  To avoid heating, the $dc$ current through the
preamplifier was kept at 1mA, enough to get an amplification
coefficient of $0.9$. The input impedance (defined by the resistance
of the sample in parallel with the load resistor and parasitic
capacitances) of the preamplifier was proportional to the
preamplifier's transfer function, $TF$ ($Vout$/$Vin$), whose real and
imaginary components were determined with the $ac$ signal applied to
the load.

Figure 1(a) shows the measured output voltage noise for a 2$\mu$m long
sample as a function of gate voltage, in the absence of any in-plane
current (zero bias) and for three different frequencies (20 kHz, 50
kHz, and 80 kHz).  The origin of this voltage is thermal noise. As
expected, the voltage noise spectral density follows the real part of
the impedance and of the transfer function, shown in Fig. 1(b).  The
background preamplifier noise, seen at small $Vg$ (small sample
resistance) in Fig.1a, has a 1/f dependence. Taking into account this
noise, about 6$nV/ \sqrt{Hz}$ for $f_0=80kHz$, we confirmed that the
measured noise is indeed thermal noise at $T=4K$ (insert in
Fig.1b). When the sample resistance is much larger than the load
resistance (at high gate voltage) the TF saturates. From the
saturation value we determined a parallel to the sample capacitance of
about 2$pF$, which is mainly the gate-drain capacitance of the
preamplifier.

The transfer function exhibits small but noticeable oscillations at
$Vg\simeq 0.5V$, as illustrated in Fig. 1(b).  Their presence suggests
that the sample is close to the mesoscopic size, where conductance is
not self-averaged but depends on a particular spatial configuration of
the fluctuation potential. In the language of percolation theory, we
can say that those oscillations reveal that the length of the sample
and the size of the critical subnetwork are comparable \cite{alex}
\cite{lee}.

\begin{figure}[h]
\begin{center}
\epsfig{figure=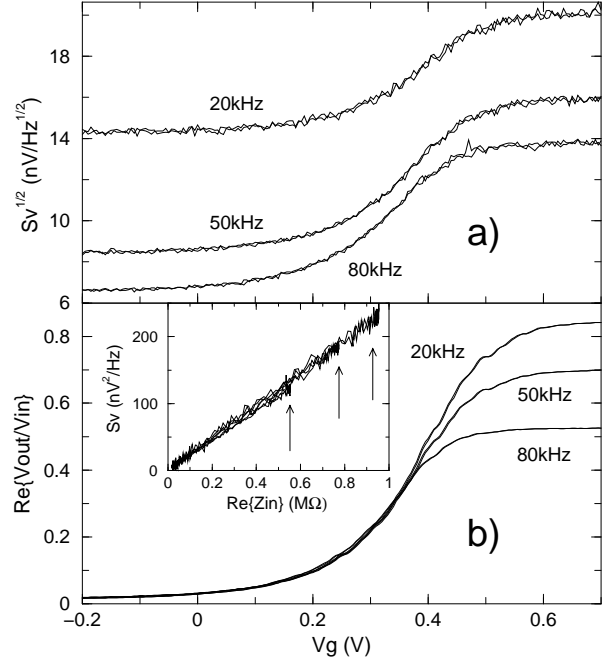,scale=0.5}
\end{center}
\caption{ The output voltage noise for a 2$\mu$m long sample (a) and
the real part of the transfer function (b) as a function of gate
voltage, at three frequencies.  The insert depicts the power spectral
density noise, obtained from (a), as a function of the input impedance
determined from (b).  Three curves are superimposed on the graph, one
for each of the three frequencies at which the voltage noise was
measured.  For each frequency, the maximum resistance at which the
voltage noise was measured is indicated by an arrow, starting with the
highest frequency on the left.  }
\end{figure}

For a fixed gate voltage, the current noise was obtained by measuring
the voltage noise spectral density as a function of the current.  The
transfer function, measured simultaneously, was then used to calculate
the current noise spectral density at the preamplifier input. The
dependence of the current noise on current at $Vg=0.5V$ is shown in
Fig.2, for $f=20$, $50$ and $80kHz$. The insignificance of the thermal
noise (the residual noise density at $Vsd=0$) is a consequence of the
fact that the sample's resistance is larger than the load resistance.
Indeed, the theoretical thermal noise is $4k_BT/(1M\Omega)\simeq
0.2\times 10^{-27}A^2/Hz$, a number that is consistent with the $I =
0$ limit of Fig. 2.

The signatures of shot noise -- linear dependence on current and
independence of frequency -- are evident in the figure.  In view of
the strongly non-linear dependence of the current on Vsd (insert in
Fig. 2), it could seem surprising that the proportionality between
noise and current is maintained in a large current range. This
proportionality suggests (as validated below) that even at the highest
voltage the in-plane electric field is still weak enough as not to
modify the critical percolation network, and indicates that the
measured shot noise is not sensitive to possible field-induced
variations of the hopping percolation paths.  From that
proportionality, a value $F = 0.59$ is obtained for the Fano factor.

\begin{figure}[t!]
\begin{center}
\epsfig{figure=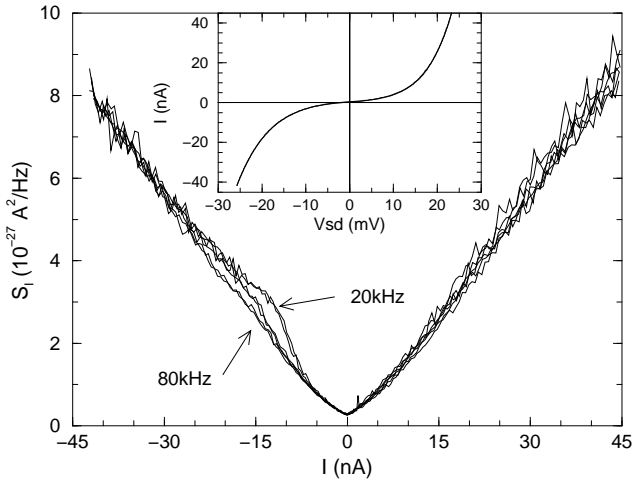,scale=0.5}
\end{center}
\caption{ Current noise spectral density as a function of the current
in a 2$\mu$m long sample subjected to a 0.5V gate voltage, showing the
proportionality between current and noise for the entire current
range.  From the slope of the curves a Fano factor F = 0.59 is
deduced.  The three superimposed curved are for measurements at the
same three frequencies of Fig.1.  The arrows point to the hump that
appears for a current of about -10 nA, which is most pronounced at the
lowest frequency.  Th insert shows the current-voltage characteristic,
measured simultaneously with the noise.  }
\end{figure}

It is noticed in Fig.2 that at $I=-10nA$ there is a hump in the noise
spectral density, which is larger at lower frequency. This is a
signature of random telegraph noise in mesoscopic structures
\cite{rts}, also seen before in hopping transport \cite{mine2}.

The Fano factor does not depend very strongly on gate voltage, as
shown in Fig.3 (top curves).  When $Vg$ decreases from $Vg=0.55V$ to
$0.2V$ $F$ drops from $0.61$ to $0.43$, which is still roughly one
half of the classical value in this range of $Vg$. Since the smaller
the gate voltage the larger the in-plane conductance, for $Vg= 0.2V$
thermal noise dominates at low current, thus the curvature in the
noise characteristic observable in Fig. 3.  The transition from
thermal to shot noise occurs above $2kT/e$.

Figure 3 also illustrates the dependence of shot noise on the length
of the current path.  Similar measurements to those on the $2\mu m$
sample are shown for a $5\mu m$ sample.  In this case, the variation
of shot noise with Vg has the same trend as before, but the change is
smaller. Most significant, however, is that shot noise is
much more suppressed in the longer sample, in which the measured Fano
factor is F = $0.2$, that is, shot noise is $1/5$ of its classical
value.

\begin{figure}[t!]
\begin{center}
\epsfig{figure=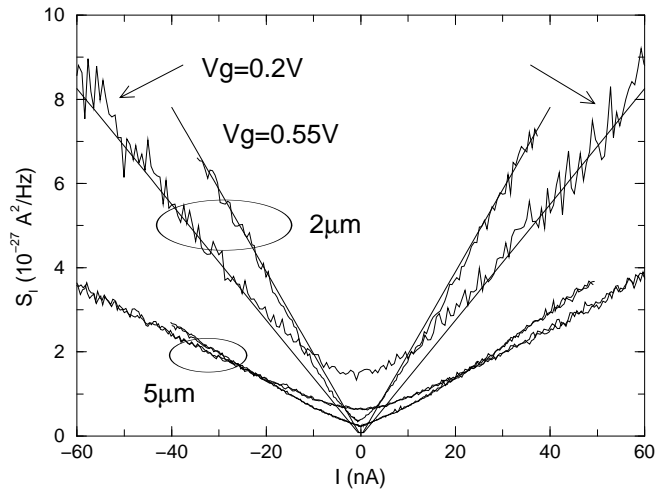,scale=0.5}
\end{center}
\caption{ Dependence of shot noise on gate voltage for samples whose
length was either $2\mu m$ or $5\mu m$.  For each set of curves
(marked by an ellipse) the gate voltages were 0.2V and 0.5V.  The
straight lines superimposed on the curves of the $2\mu m$ set give
Fano factors of $0.61$ and $0.43$ for $Vg=0.55V$ and $0.2V$,
respectively.  }
\end{figure}

We can explain our results if we assume that there is a characteristic
scaling length $L_0\simeq 1\mu m$ in 2D hopping for both samples, such
that the Fano factor is just the ratio of that length to the length of
the sample, $F=L_0/L$. It is reasonable to assume also that this scale
is a characteristic of the homogeneity of the sample, which is the
distance between hard hops of the critical percolation
subnetwork. Then, the trend for noise suppression on $Vg$ reflects the
fact that hopping becomes more uniform as the sample is driven towards
the insulator-metal transition.

To justify these assumptions we can take into account the fact that
when an electric field is applied, only hard hops along the field
direction are modified (e.g., decrease their resistances) and thus the
network is separated into a set of equivalent parallel chains.  In
this case, the total shot noise would be that of a single chain,
which, as we have seen, will have a Fano factor inversely proportional
to the number of hard hops in the chain.

The distance between hard hops in the percolation subnetwork can
be obtained using \cite{shklovski}: 
\begin{equation}
L_0=l(T)\Bigl(\frac{T_0}{T}\Bigr)^{\frac{\nu}{d+1}},
\end{equation}
where $l(T)$ is the characteristic hopping length in the zero-field
limit, $T_0$ is a characteristic temperature inversely proportional to
the density of states at the Fermi level and to the localization
radius $a$, $d$ is the effective dimensionality of the system (d=$1$
for hopping with Coulomb gap and d=$2$ for 2D hopping), and $\nu$ is
the critical index of the correlation radius, which is about 1.3 for a
2D system. In turn, the hopping length can be estimated within the
percolation model, in which the non-linear conductance G(E,T) is
written \cite{hill,pollak,shklovski} as
\begin{equation}
\ G(E,T)=\frac{I}{Vsd}=G(0,T)exp\Bigl(\frac{eEl(T)}{k_BT}\Bigr),
\end{equation}
where $E$ is the electric field. (The other symbols have their usual
meaning.)  This expression is only valid in the low-field regime, that
is, when $eEa<k_BT$.  The hopping length depends on temperature as
\begin{equation}
l(T)=a\Bigl(\frac{T_0}{T}\Bigr)^{\frac{1}{1+d}}.
\end{equation}

When Eq. (2) was used in combination with the experimental $I-Vsd$
characteristics (Fig. 2, insert) a value of $l\simeq 0.08\mu m$ was
obtained for the hopping length of the $2\mu m$ sample at T = 4K. The
localization radius was estimated to be 100-130\AA\, from Eq. (3) and
the experimental temperature dependence of the zero-field conductance
in the range $2K<T<30K$.  The 30\AA\ variance of $a$ reflects the
difficulty in discerning experimentally between $1/2$ and $1/3$ for
the exponent in Eq. (3).  This estimation of the localization radius
is consistent with a fluctuation potential created by interface
impurities separated about 200\AA\ from each other (surface density of
$2\times 10^{11}cm^{-2}$ \cite{emel}) and validates our low-field
assumption since for $Vsd=30mV$ (insert, Fig. 2) it is $Ea\simeq
0.15meV<k_BT/e\simeq 0.36meV$.

Finally, from Eq.(1) we get either $L_0\simeq 0.8\mu m$ or $\simeq
1.2\mu m$, depending, again, on whether we use $d=1$ or $2$.  A
similar analysis for the data on the $5\mu m$ sample yielded a value
of 1$\mu m$ for the inter-node distance.  This result tells us that if
two nodes were separated by 1$\mu m$, than there would be two hard
hops in a $2\mu m$ sample and five such hops in a $5\mu m$ sample.
Consequently, according to the above discussion, the corresponding
Fano factor in the shot noise formula should be $0.5$ and $0.2$,
respectively.  These values are in excellent agreement with the
experimental results.

Although these results have been obtained for a 2D hole gas, they
should be general to any other system in the hopping regime. It also
follows from our results that by decreasing the sample length even
further one could obtain full shot noise, corresponding to tunneling
through only one hard hop along the current direction. Interestingly,
a further decrease in length and in temperature should cause a
transition to resonant tunneling transport, such as resonant tunneling
through impurities \cite{mine3}. The only calculation available in
such a regime for the shot noise is for tunneling through one
impurity, done recently \cite{nazarov}, in which $F=3/4$. We hope that
the experiments presented here will stimulate calculations in the
entire hopping conduction regime.

We are thankful to Alexander Korotkov for numerous discussions. This
work has been sponsored by the Department of Energy (DOE Grant
No. DE-FG02-95ER14575).

\end{multicols}

\end{document}